\documentclass[graybox]{svmult}

\usepackage{mathptmx}       
\usepackage{helvet}         
\usepackage{courier}        
\usepackage{makeidx}         
\usepackage{graphicx}        
\usepackage{multicol}        
\usepackage[bottom]{footmisc}

\makeindex             

\usepackage{amsfonts}
\usepackage{amssymb}
\usepackage{amsmath}
\usepackage{color, soul}
\usepackage{algpseudocode}
\usepackage{algorithm}
\usepackage{breqn}
\usepackage[numbers]{natbib}
\usepackage{url}
\usepackage[colorlinks=true]{hyperref}
\usepackage{mathrsfs}
 
\newcommand{\anomalyset}{\mathfrak{K}}
\newcommand{\indexset}{\mathfrak{N}}
\newcommand{\calD}{\mathcal{D}}
\newcommand{\calN}{\mathcal{N}}
\newcommand{\R}{\mathbb{R}}
\renewcommand{\E}{\mathbb{E}}
\newcommand{\tpos}{\mathsf{T}}
\newcommand{\xdott}{{x}_{(\centerdot, t)}}

\begin{document}

\title*{Compressed Anomaly Detection with Multiple Mixed Observations}
\author{Natalie~Durgin, Rachel~Grotheer, Chenxi~Huang, Shuang~Li, Anna~Ma, Deanna~Needell  and Jing~Qin}

\institute{Natalie Durgin \at Spiceworks \\ Austin, TX, 78746.\\ \email{njdurgin@gmail.com}
\and Rachel Grotheer \at Goucher College \\ Baltimore, MD 21204.
\and Chenxi Huang \at Yale University \\ New Haven, CT 06511. \\  \email{chenxi.huang@yale.edu}
\and Shuang Li \at Colorado School of Mines \\ Golden, CO 80401. 
\and Anna Ma \at Claremont Graduate University \\ Claremont, CA 91711. 
\and Deanna Needell \at University of California, Los Angeles \\ Los Angeles, CA 90095. 
\and Jing Qin \at Montana State University \\ Bozeman, MT 59717. }
\authorrunning{Durgin, Grotheer, Huang, Li, Ma, Needell, Qin}
%
%
\maketitle

\abstract{We consider a collection of independent random variables that are identically distributed, except for a small subset which follows a different, anomalous distribution. We study the problem of detecting which random variables in the collection are governed by the anomalous distribution. Recent work proposes to solve this problem by conducting hypothesis tests based on mixed observations (e.g. linear combinations) of the random variables. Recognizing the connection between taking mixed observations and compressed sensing, we view the problem as recovering the ``support'' (index set) of the anomalous random variables from multiple measurement vectors (MMVs). Many algorithms have been developed for recovering jointly sparse signals and their support from MMVs. We establish the theoretical and empirical effectiveness of these algorithms in detecting anomalies. We also extend the LASSO algorithm to an MMV version for our purpose. Further, we perform experiments on synthetic data, consisting of samples from the random variables, to explore the trade-off between the number of mixed observations per sample and the number of samples required to detect anomalies.}

\section{Introduction}
\label{sec:intro}
The problem of anomaly detection has been the focus of interest in many fields of science and engineering, including network tomography, cognitive radio, and radar \cite{poor2009quickest,xia2006inference,lai2008quickest,basseville1993detection}. In this paper, we study the problem of identifying a small number of anomalously distributed random variables within a much larger collection of independent and otherwise identically distributed random variables. We call the random variables following the anomalous distribution \textit{anomalous random variables}.  A conventional approach to detecting these anomalous random variables is to sample from each random variable individually and then apply hypothesis testing techniques~\cite{MN11,MN11b,MTN12,MTN12}. 

A recent paper~\cite{CXL16} proposes to perform hypothesis testing on mixed observations (e.g. linear combinations) of random variables instead of on samples from individual random variables. They call this technique \textit{compressed hypothesis testing}. Such an approach is motivated by the recent development of \textit{compressed sensing} \cite{CandesCS,DonohoCS,FR12:Mathematical-Introduction,eldar2012compressed}, a signal processing paradigm that shows a small number of random linear measurements of a signal is sufficient for accurate reconstruction. Now a large body of work in this area shows that optimization-based \cite{DH01:Uncertainty-Principles,CT05:Decoding,Don06:Compressed-Sensing,CRT06:Stable,Tro04:Just-RelaxArt} and iterative \cite{TG07:Signal-Recovery,NV07:ROMP-Stable,BD08:Iterative} methods can reconstruct the signal accurately and efficiently when the samples are taken via a sensing matrix satisfying certain incoherence properties \cite{CT05:Decoding,CT06:Near-Optimal}. Compressed sensing is also studied in a Bayesian framework, where signals are assumed to obey some prior distribution \cite{ji2008bayesian,yu2011statistical,baron2010bayesian}.

The results presented in \cite{CXL16} show that the ``mixed'' measurement approach achieves better detection accuracy from fewer samples when compared to the conventional ``un-mixed'' approach. However, compressed hypothesis testing requires that the distributions of the random variables are known \textit{a priori}, which may not be available in practice. Further, as the authors pointed out, their proposed approach requires conducting a large number of hypothesis tests, especially when the number of random variables in the collection is large, rendering such an approach computationally prohibitive. Two efficient algorithms are proposed as alternatives in \cite{CXL16}, but no analytical study of their performance is provided.

We propose new methods for detecting anomalous random variables that require minimal knowledge of the distributions, are computationally efficient, and whose performance is easy to characterize. We begin by generalizing the compressed hypothesis testing method and posing our problem as a \textit{multiple measurement vector} (MMV) problem \cite{HN05,BWDSB05,DWBSB13,Duarte2006IPSN,chen2006trs,cotter2005ssl,Mishali08rembo,Berg09jrmm}.
In the MMV compressed sensing setting, a collection of signals are recovered simultaneously, under the assumption that they have some commonalities, such as sharing the same support. A related vein of work involves signals that are smoothly varying, where the support may not be consistent but changes slowly over time \cite{angelosante2009compressed,filos2013tracking,Patterson2014}. 
While the compressed hypothesis testing in \cite{CXL16} is certainly motivated by compressed sensing techniques, the authors do not formally frame the anomaly detection problem in the compressed sensing setting. Also, they do not focus on compressed sensing algorithms that might eliminate the need for prior knowledge of the distributions, and might lead to more efficient detection for large collections of random variables.

In the following, we view the collection of random variables as a random vector and aim to identify the indices of the anomalous random variables within the random vector. We also draw an analogy between the collection of independent samples from the random vector and an ensemble of signals where in practice these signals often become available over time.  More specifically, we consider a random vector, $X = (X_1, \ldots, X_N)$, where the $X_n$'s are independent random variables. We assume that each $X_n$ follows one of two distributions, $\calD_1, \calD_2$. We call $\calD_1$ the \textit{prevalent distribution}, and $\calD_2$ the \textit{anomalous distribution}. We let $\indexset=\{n\in\mathbb{N}: 1\leq n \leq N\}$ denote the index set of the random variables, $X_n$, and let $\anomalyset$ denote the index set of the $K$ random variables that follow the anomalous distribution. Let $\xdott \in \R^N$ denote the independent realization of the random vector at time $t$.  At each time-step $t$, we obtain $M$ mixed observations by applying the sensing matrix $\phi_t\in \R^{M\times N}$,
\[
y_t=  \phi_t \xdott,  \text{ } 1\leq t\leq T,
\]
with $y_t\in \mathbb{R}^{M}$. Thus the goal of the anomaly detection problem in this setting is to recover the index set $\anomalyset$ from the MMVs $y_t$, $t=1,\cdots,T$.

The signals $\xdott$ in our formulation are not necessarily sparse and may have different supports since they are samples from a random vector and are changing over time. Nevertheless, there is still a close connection between our formulation and that for recovering the common sparse support of a collection of signals from MMVs. The index set of the anomalous random variables, which corresponds to the index set of the anomalies (realizations of anomalous random variables) in the signals $\xdott$, is shared by all signals. This index set can thus be viewed as the common ``support'' of the anomalies in the signals, which motivates us to consider the applicability of many MMV algorithms designed for signal reconstruction. Further, the analytical studies of many of these algorithms are readily available. We therefore investigate which of these MMV algorithms can be applied or adapted to the anomaly detection problem under consideration and analyze their performance in detection accuracy in theory and through numerical experiments. We focus on algorithms presented in~\cite{BWDSB05}. 

\subsection{Contributions} 
In this paper, by extending the definitions of two so-called \textit{joint sparsity models} (JSMs) from~\cite{BWDSB05}, we introduce two new signal models, JSM-2R and JSM-3R, for the problem of anomaly detection. For JSM-2R and JSM-3R signals, we adapt several MMV signal reconstruction algorithms to anomaly detection. Additionally, we develop a new algorithm for the JSM-2R model that extends the \textit{Least Absolute Shrinkage and Selection Operator} (LASSO) algorithm~\cite{CDS01} to the MMV framework. We show theoretically and numerically that these algorithms accurately detect the anomalous random variables. We also provide numerical results which demonstrate the trade-off between the number of time-steps, and the number of mixed observations per time-step needed to detect the anomalies. 

\subsection{Organization}
In Section~\ref{sec:method}, we introduce the models JSM-2R, JSM-3R and the four algorithms we have repurposed from MMV signal recovery into MMV anomaly detection, as well as our new LASSO algorithm. We also provide theoretical guarantees in this section. In Section~\ref{sec:experiments}, we explore the performance of these algorithms by conducting numerical experiments for some strategic choices of the parameters involved. Finally, we conclude in Section~\ref{sec:conclusion}. To help keep track of notation, we provide a handy reference table in Section~\ref{sec:notation}. We adopt the convention that random variables will be upper case and their realizations will be lower case. All matrix entries will have two, subscripted indices. The first index will indicate the row position, the second will indicate the column position.

\section{Method}
\label{sec:method}
In this section, we introduce two new signal models for the anomaly detection problem and describe five algorithms for detecting anomalous random variables under these signal models. We also provide theoretical guarantees for the algorithms. 

Recall that we consider the problem of detecting $K$ anomalous random variables from a collection of $N$ random variables where $K\ll N$. The anomalous random variables have a different probability distribution from that of the remaining $N-K$ random variables. We seek to identify the $K$ anomalous random variables, from $T$ independent realizations of the $N$ random variables. To emphasize our framing of this random variable problem as a compressed sensing problem, we refer to the independent realizations as signals. These $T$ signals have an important commonality: they share the same indices of anomalous entries (realizations of anomalous random variables).

Commonality among signals has already been explored in the field of distributed compressed sensing for recovering signals that have specific correlation among them. Three \textit{joint sparsity models} (JSMs) were introduced in \cite{BWDSB05} to characterize different correlation structures.  To utilize the commonality of the signals for anomaly detection, we propose two new signal models that are motivated by two of the JSMs defined in \cite{BWDSB05}, namely, JSM-2 and JSM-3. Since the signals under consideration are realizations of random variables, we term the new models JSM-2R and JSM-3R, respectively, where the appended ``R'' indicates the ``random variable'' version of the existing JSMs. 

Before we define the new models, We first briefly describe JSM-2 and JSM-3. The JSM-2 signals are jointly sparse signals that share the same support (the indices of non-zero entries). The JSM-3 signals consist of two components: a non-sparse ``common component'' shared by all signals and a sparse ``innovation component'' that is different for each signal. But the innovation components of the JSM-3 signals share the same support. We next extend these definitions to the signals in the anomaly detection setting. The new JSM-2R and JSM-3R models are defined as follows:

\begin{definition}[JSM-2R and JSM-3R]
\label{def:jsm2r3r}
Let the random variable $X_n\sim\calD_1$ if $n\notin \anomalyset$ and $X_n\sim\calD_2$ if $n\in \anomalyset$ where $\anomalyset$ is the set of the anomalous indices. For a signal ensemble $x\in \R^{N\times T}$ where each of its entries $x_{(n,t)}$ denotes the realization of $X_n$ at time $t$, 
\begin{enumerate}
\item $x$ is a JSM-2R signal ensemble when: $|x_{(n,t)}|$ is small if $n\notin\anomalyset$ and $|x_{(n,t)}|$ is large if $n\in\anomalyset$;
\item $x$ is a JSM-3R signal ensemble when: $x_{(n,t)}=x_n^C+x_{(n,t)}^I$ such that $|x_{(n,t)}^I|$ is small if $n\notin\anomalyset$ and $|x_{(n,t)}^I|$ is large if $n\in\anomalyset$. $x_n^C$ is a common component shared by all $t$, and $x_{(n,t)}^I$ is an innovation component that is different for different $t$.
\end{enumerate}
\end{definition}

The JSM-2R signal model assumes a small amplitude for variables generated from the prevalent distribution and a large amplitude for variables generated from the anomalous distribution. Such a model characterizes a scenario where anomalies exhibit large spikes. This model relates to a sparse signal model where the support of the sparse signal corresponds to the set of indices of the anomalous random variables. In fact, when $\calD_1=\calN(0,\sigma^2)$ and $\calD_2=\calN(\mu,\sigma^2)$ with $\mu\neq 0$, the JSM-2R signal is a sparse signal with additive Gaussian noise. An example of anomalies following the JSM-2R model is a network where some of the sensors completely malfunction and produce signals with vastly different amplitudes than the rest of the sensors.

\begin{figure}[h!]
\includegraphics[width=4.25in]{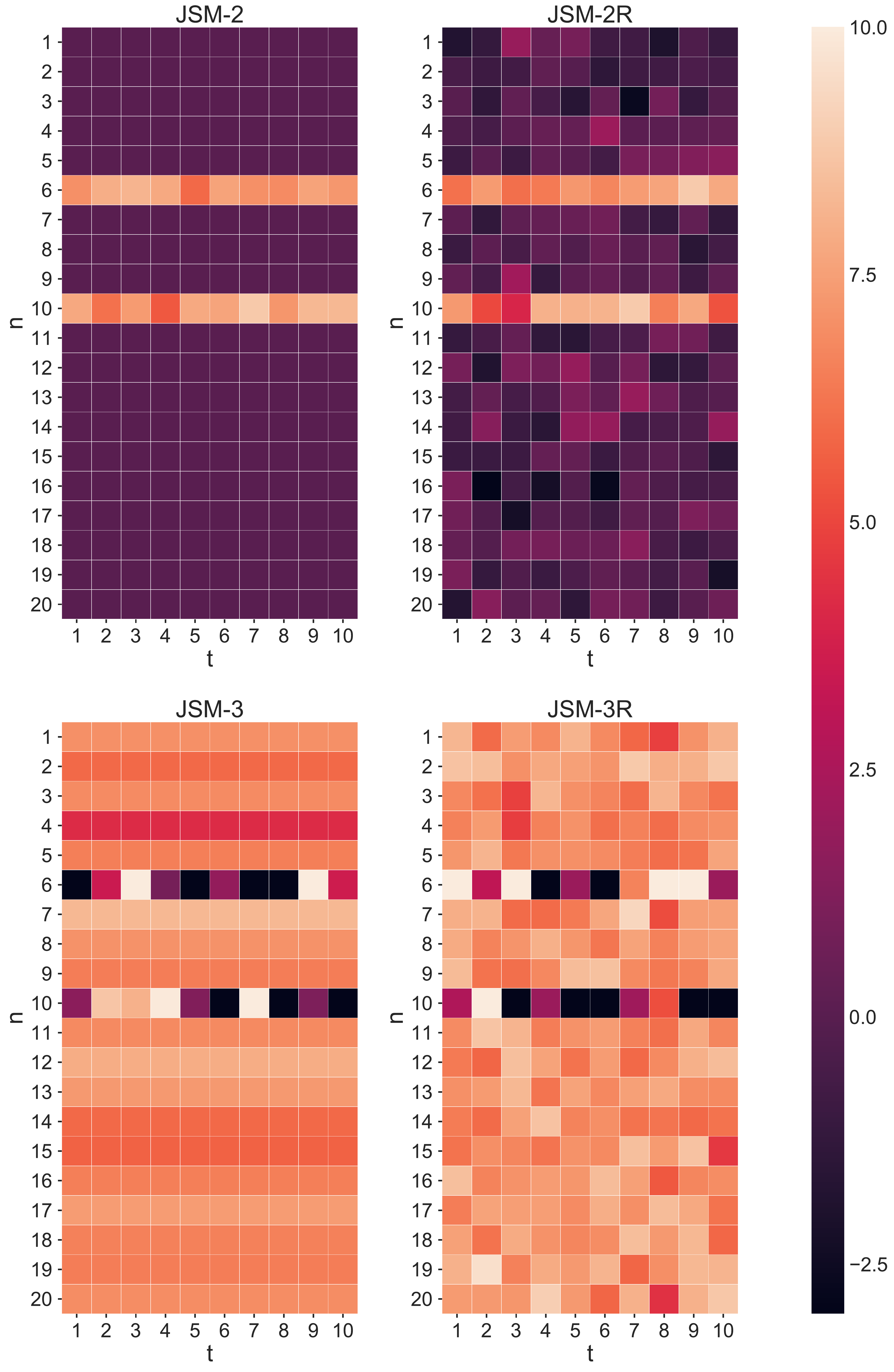}
\caption[Joint Sparsity Model Examples]{Depiction of the existing joint sparsity models (JSM-2 and JSM-3) and the new models developed for anomaly detection (JSM-2R and JSM-3R). The distributions used to generate this example are the same as the ones used for the numerical experiments in Section~\ref{sec:experiments}, see Table~\ref{tab:exp}. The index set of the anomalies is $\anomalyset=\{6, 10\}$.}
\label{fig:jsmex}
\end{figure}

Different from the JSM-2R signals, the JSM-3R signal model introduced above does not have constraints on the amplitude of the signal entries $x_{(n,t)}$. Rather the signals at different time-steps are assumed to share an unknown common component $x_n^C$ while having a different innovation component $x_{(n,t)}^I$ for signals at different time-steps. Of note, the common component $x_n^C$ from the prevalent distribution may or may not be the same as that from the anomalous distribution. Further, the innovation component $x_{(n,t)}^I$ is assumed to follow the JSM-2R signal model. Such a model characterizes a scenario where there exists a background signal that does not change over time and the anomalies exhibit large spikes on top of the background signal. Because of the common component, the JSM-3R signals no longer correspond to a sparse signal model. The JSM-3R model has applications in geophysical monitoring where a constant background signal is present and anomalies appear as large spikes of erratic behavior. Figure~\ref{fig:jsmex} provides a visual illustration of the model nuances. 

\subsection{Algorithms}
We briefly describe the five algorithms we study in this paper, among which three are for JSM-2R signals and two are for JSM-3R signals. Two of the algorithms for JSM-2R signals were originally proposed for recovering JSM-2 signals, including the \textit{one-step greedy algorithm} (OSGA) and the \textit{multiple measurement vector simultaneous orthogonal matching pursuit} (MMV-SOMP) algorithm. We further propose a new MMV version of the LASSO algorithm for detecting anomalies for JSM-2R signals and investigate its performance via numerical experiments. The two algorithms for JSM-3R were also proposed in \cite{BWDSB05} for recovering JSM-3 signals, including the \textit{Transpose Estimation of Common Component} (TECC) algorithm and the \textit{Alternating Common and Innovation Estimation} (ACIE) algorithm. 

For each of the presented algorithms, the goal is to identify the indices of the anomalous random variables from the mixed measurements $y_t=\phi_t \xdott$ for $t~=~1,~2,~\ldots,~T.$ The number of anomalies $K$ is assumed to be known \textit{a priori}. We first describe three algorithms we applied to anomaly detection for JSM-2R signals.

\subsubsection{OSGA}
The OSGA algorithm is a non-iterative greedy algorithm introduced in \cite{BWDSB05} to recover the support of JSM-2 signals based on inner products of the measurement and columns of the sensing matrix (Algorithm \ref{alg:osga}). We show in Theorem \ref{thm:osga} that with some condition on the prevalent and anomalous distributions, the OSGA algorithm is able to recover the anomaly indices under the JSM-2R model, using a small number of measurements per time-step. Although the OSGA algorithm is shown to work asymptotically, it may not perform well when only a small number of time-steps are available. Empirical evidence has confirmed this conjecture when the OSGA algorithm is used to reconstruct JSM-2 signals \cite{BWDSB05}. Thus we further consider approaches like matching pursuit \cite{MZ93,PRK93} for our problem.  Next, we describe the MMV version of orthogonal matching pursuit algorithm proposed in \cite{BWDSB05}.

\begin{algorithm}[h]
\caption[OSGA]{OSGA}
\label{alg:osga}
\begin{algorithmic}[1]
\State\textbf{Input:} {$y_1, \ldots, y_T$, $\phi_t$, $K$}.
\State\textbf{Output:} $\widehat{\anomalyset}$.
\For{$n=1,2,\ldots,N$}
\State Compute $\xi_n=\frac{1}{T}\sum_{t=1}^{T} {\langle y_t,\phi_{t(\centerdot,n)}\rangle }^2$
\EndFor
\State \textbf{return} $\widehat{\anomalyset}=\{n$, for the $K$ largest $\xi_n\}$
\end{algorithmic}
\end{algorithm}

\subsubsection{MMV-SOMP} The MMV-SOMP algorithm is an iterative greedy pursuit algorithm for recovery of jointly sparse signals. SOMP was first proposed in \cite{TGS05}  and was adapted to the MMV framework in \cite{BWDSB05}. Since our focus is not on signal recovery but on detecting anomalous entries, we adapt this algorithm for our JSM-2R signal model. The adapted algorithm is presented in Algorithm \ref{alg:mmvsomp}, which identifies the anomaly indices one at a time. In each iteration, the column index of the sensing matrices that accounts for the largest residual across signals of all time-steps is selected. Then the remaining columns of each sensing matrix (for each time-step) are orthogonalized. The algorithm stops after $K$ iterations where $K$ is the number of anomalous random variables. We show through numerical experiments in Section \ref{sec:experiments} that the adapted MMV-SOMP algorithm performs better than the OSGA algorithm for a small number of time-steps.

\begin{algorithm}[h!]
\caption[MMV-SOMP]{MMV-SOMP}
\label{alg:mmvsomp}
\begin{algorithmic}[1]
\State\textbf{Input:} {$y_1, \ldots, y_T$, $\phi_t$, $K$}.
\State\textbf{Output:} $\widehat{\anomalyset}$.
\State\textbf{Initialize:} $\widehat{\anomalyset}=\emptyset$, residual $r_t^0=y_t$.
\For{$k=1,\ldots,K$}.
\State Select $$n_k= \underset{n}{\operatorname{arg\,max}} \sum_{t=1}^T\frac{|\langle r_t^{k-1},\phi_{t(\centerdot, n)}\rangle |}{{\|\phi_{t(\centerdot,n)}\|}_2}.$$
\State Update $\widehat{\anomalyset}=[\widehat{\anomalyset},n_k]$.
\State Orthogonalize selected basis vector against previously selected vectors for~all~$t$, $1\leq~t\leq~T$:
\begin{align*}
\gamma_t^0&=\phi_{t(\centerdot,k)}, &\text{if $k=1$,} \\
\gamma_t^k&=\phi_{t(\centerdot, n_k)}-\sum_{l=0}^{k-1} \frac{\langle\phi_{t(\centerdot, n_k)},\gamma_t^l\rangle }{\|\gamma_t^l\|_2^2}\gamma_t^l, &\text{if $k>1$.} \\
\end{align*}
\State Update the residual for all $t$, $1\leq t\leq T$,  $$r_t^k=r_t^{k-1}-\frac{\langle r_t^{k-1},\gamma_t^{k}\rangle }{\|\gamma_t^{k}\|_2^2}\gamma_t^{k}.$$
\EndFor
\State \textbf{return} $\widehat{\anomalyset}$
\end{algorithmic}
\end{algorithm}

\subsubsection{MMV-LASSO} The LASSO algorithm aims to find a sparse solution to the regression problem by constraining the $L_1$ norm of the solution~\cite{CDS01}. The LASSO algorithm was also considered in \cite{CXL16} as an efficient algorithm for anomaly detection from mixed observations. However, the authors of \cite{CXL16} considered the LASSO algorithm when using only one measurement at each time-step. In this paper, we further extend the LASSO algorithm to a more general setting for MMV and term it the MMV-LASSO algorithm. The MMV-LASSO algorithm is described in Algorithm \ref{alg:mmvlasso}. The measurements $y_t\in \R^M$ up to $T$ time-steps are concatenated vertically to become a vector $y\in \R^{(MT)\times 1}$; the sensing matrices $\phi_t\in\R^{M\times N}$ are also concatenated vertically to become $\phi\in\R^{(MT)\times N}$. The concatenated measurements and sensing matrices are then fed to the regular LASSO algorithm, where the anomaly indices are found by taking indices corresponding to the $K$ largest amplitudes of the estimate. The LASSO problem, that is, Step 4 in Algorithm 3, can be tackled by various approaches \cite{EHJT04,KS07}, which is out of scope of this paper. 

\begin{algorithm}[h!]
\caption[MMV-LASSO]{MMV-LASSO}
\label{alg:mmvlasso}
\begin{algorithmic}[1]
\State\textbf{Input:} {$y_1, \ldots, y_T$, $\phi_t$, $K$}.
\State\textbf{Output:} $\widehat{\anomalyset}$.
\State Let $y=[y_1^{\tpos},\ldots,y_T^{\tpos}]^{\tpos}$ and $\phi=[\phi_1^{\tpos},\ldots,\phi_T^{\tpos}]^{\tpos}$
\State Solve $$\hat{x}=\underset{x}{\operatorname{arg\,min}}\frac{1}{2}\|y-\phi x\|_2^2+\lambda\|x\|_1 $$
\State Let $\hat{x}_n$ denote the $n$-th element of $\hat{x}$
\State \textbf{return} $\widehat{\anomalyset}=\{n, $ for the $K$ largest $|\hat{x}_n|\}$
\end{algorithmic}
\end{algorithm}

We next describe two algorithms for anomaly detection for JSM-3R signals.

\subsubsection{TECC} 
The key difference between JSM-2R and JSM-3R signals is that JSM-3R signals share a common component that is unknown. Thus the two algorithms for the JSM-3R signals aim to first estimate the common component from the mixed measurement and subtract the contribution of this component from the measurement. The TECC algorithm was proposed in \cite{BWDSB05} for recovering JSM-3 signals. We also adapt the algorithm to focus only on detecting the anomalous indices of JSM-3R signals, and the adapted algorithm can be found in Algorithm \ref{alg:tecc}. The first step of the TECC algorithm estimates the common component of the JSM-3R signals. Using this estimate, the contribution of the remaining innovation component to the measurement can be estimated. Then algorithms for JSM-2R signals can be applied to identify the anomaly indices. We show in Theorem \ref{thm:mmvtecc} that the TECC algorithm is able to identify the anomalous variables under some conditions on the prevalent and anomalous distributions. Similar to the OSGA algorithm, while Theorem~\ref{thm:mmvtecc} guarantees the success of the TECC algorithm in the asymptotic case as $T$ goes to infinity, it may not perform well for a small $T$. Next we describe an alternative algorithm also proposed in \cite{BWDSB05} for cases with a small $T$.

\begin{algorithm}[h]
\caption[TECC]{TECC}
\label{alg:tecc}
\begin{algorithmic}[1]
\State\textbf{Input:} {$y_1, \ldots, y_T$, $\phi_t$, $K$}.
\State\textbf{Output:} $\widehat{\anomalyset}$.
\State Let $y=[y_1^{\tpos},\ldots,y_T^{\tpos}]^{\tpos}$, and $\phi=[\phi_1^{\tpos},\ldots,\phi_T^{\tpos}]^{\tpos}$
\State Calculate $\widehat{x^C}=\frac{1}{TM}\phi^{\tpos}y$.
\State Calculate $\widehat{y}_t=y_t-\phi_t\widehat{x^C}$.
\State Estimate $\widehat{\anomalyset}$ from $\widehat{y}_t$ by Algorithm \ref{alg:osga}, \ref{alg:mmvsomp} or \ref{alg:mmvlasso} 
\State \textbf{return} $\widehat{\anomalyset}$
\end{algorithmic}
\end{algorithm}

\subsubsection{ACIE} 
The ACIE algorithm is an extension of the TECC algorithm, also introduced in \cite{BWDSB05}, based on the observation that the initial estimate of the common component may not be sufficiently accurate for subsequent steps. Instead of one-time estimation in the TECC algorithm, the ACIE algorithm iteratively refines the estimates of the common component and the innovation components. The ACIE algorithm can also be easily adapted for the JSM-3R signals for anomaly detection. In the ACIE algorithm described in Algorithm \ref{alg:acie}, we first obtain an initial estimate of the anomaly index set $\widehat{\anomalyset}$ using the TECC algorithm. Then for each iteration, we build a basis $B_t$ for $\R^M$ where $M$ is the number of measurements at each time-step: $B_t=[\phi_{t,\widehat{\anomalyset}}, q_t]$, where $\phi_{t,\widehat{\anomalyset}}$ is the subset of the basis vectors in $\phi_t$ corresponding to the indices in $\widehat{\anomalyset}$ and $q_t$ has orthonormal columns that spans the orthogonal complement of $\phi_{t,\widehat{\anomalyset}}$. Then we can project the measurements onto $q_t$ to obtain the part of the measurement caused by signals not in $\anomalyset$:

\begin{align}
\widetilde{y}_t&={q_t}^{\tpos}y_t,\label{eq:acie1}\\
\widetilde{\phi}_t&={q_t}^{\tpos}\phi_t\label{eq:acie2}.
\end{align}
Then $\widetilde{y}_t$ and $\widetilde{\phi}_t$ are used to refine the estimate of the common component. After subtracting the contribution of this estimated common component, algorithms such as OSGA and MMV-SOMP described above can be applied to detect the anomalies.

\begin{algorithm}[h]
\caption[ACIE]{ACIE}
\label{alg:acie}
\begin{algorithmic}[1]
\State\textbf{Input:} {$y_1, \ldots, y_T$, $\phi_t$, $K$, $L$ (iteration counter)}.
\State\textbf{Output:} $\widehat{\anomalyset}$.
\State Let $y=[y_1^{\tpos},\ldots,y_T^{\tpos}]^{\tpos}$
\State Obtain an initial estimate of $\widehat{\anomalyset}$ from Algorithm \ref{alg:tecc}
\For{$l=1,2,\ldots,L$}
\State Update $\widetilde{y}_t$ and $\widetilde{\phi}_t$ according to Equations (\ref{eq:acie1}) and (\ref{eq:acie2}) for all $t$, $1\leq t\leq T$
\State Update $\widetilde{x^C}={\widetilde{\phi}}^{\dagger}\widetilde{y}$, where $\widetilde{y}=[\widetilde{y}_1^{\tpos},\cdots, \widetilde{y}_T^{\tpos}]^{\tpos}$, $\widetilde{\phi}=[{\widetilde{\phi}_1}^{\tpos},\ldots,{\widetilde{\phi}_T}^{\tpos}]$ and $\widetilde{\phi}^{\dagger}={(\widetilde{\phi}^{\tpos}\widetilde{\phi})}^{-1}\widetilde{\phi}^{\tpos}$
\EndFor
\State Calculate $\widehat{y}_t=y_t-\phi_t\widetilde{x^C}$
\State Estimate $\widehat{\anomalyset}$ from $\widehat{y}_t$ by Algorithm \ref{alg:osga}, \ref{alg:mmvsomp} or \ref{alg:mmvlasso} 
\State \textbf{return} $\widehat{\anomalyset}$
\end{algorithmic}
\end{algorithm}
\subsection{Theoretical Guarantees}
In this section we show theoretically that Algorithm~\ref{alg:osga} and Algorithm~\ref{alg:tecc} (coupled with Algorithm~\ref{alg:osga} in step 6) can detect anomalies for the JSM-2R and JSM-3R settings, respectively. 

Recall that Algorithm \ref{alg:osga} is designed for JSM-2R signals where variables generated from the prevalent distribution are much smaller in amplitude than those from the anomalous distribution. The following theorem shows that for JSM-2R signals, the OSGA algorithm is able to identify the indices of the anomalous variables asymptotically, with very few measurements at each time-step. 

\begin{theorem}\label{thm:osga}[Adapted from~\cite{BWDSB05} Theorem 8]
Let the $M\times N$ sensing matrix, $\phi_t$, contain entries that are i.i.d $\sim\calN(0,1)$ at each time-step $t$. Suppose the random variables, $X_n$, are distributed with $\calD_1=\mathcal{N}(0,\sigma_1^2)$ if $n\notin \anomalyset$ and $\calD_2=\mathcal{N}(\mu_2,\sigma_2^2)$ if $n\in \anomalyset$. Assuming $\mu_2^2+\sigma_2^2>\sigma_1^2$, then with $M\geq 1$ measurements per time-step, OSGA recovers $\anomalyset$ with probability approaching one as $T\rightarrow\infty$. 
\end{theorem}

Before diving into the proof of Theorem~\ref{thm:osga}, we first observe that the signals correspond to the JSM-2R signals: with a zero mean and a potentially small variance $\sigma_1^2$ for the prevalent distribution $\calD_1$, the signal entry $x_{(n,t)}$, $n\notin\anomalyset$ (i.e. the realization of $X_n$ at the time-step $t$) is expected to have small amplitude. In contrast, with a non-zero mean $\mu_2$ and a similar or possibly larger variance $\sigma_2^2$ for the anomalous distribution $\calD_2$, the amplitude of $x_{(n,t)},  n\in\anomalyset$ is expected to be much larger.

\begin{proof}
\smartqed
We assume, for convenience and without loss of generality, that the anomalous random variables are indexed by, $\anomalyset=\{1,2,\ldots,K\}$, and the prevalent random variables are indexed by $\indexset\backslash \anomalyset = \{K+1, \ldots, N\}$. Consider that the test statistic $\xi_n = \frac{1}{T} \sum_{t=1}^T \langle y_t, \phi_{t(\centerdot, n)}\rangle^2$ is the sample mean of the random variable $ \langle Y, \Phi_{(\centerdot, n)}\rangle^2$, so by the Law of Large Numbers, 
 
$$\lim_{T\rightarrow \infty} \xi_n = \E[\langle Y, \Phi_{(\centerdot, n)}\rangle^2].$$ 

We select an arbitrary index $n$ from each of the anomalous random variable index set and the prevalent random variable index set, and compute $\E[{\langle Y,\Phi_{(\centerdot, n)}\rangle}^2]$ in each case. As the final step, we compare the expected values of the two $\xi_n$ and establish that they are distinguishable under very general conditions. Without loss of generality, we select $n=K+1$  for the ``prevalent case'' and  $n=1$ for the ``anomalous case''. Note that \cite{BWDSB05} refers to these cases respectively as the ``bad statistics'' and the ``good statistics'' in their setting. For them, ``bad'' reflects an incorrect estimate of the sparse support and ``good'' reflects a correct estimate of the sparse support. 

\runinhead{Prevalent Case:}
Substituting $\Phi X$ for $Y$ in  $\langle Y, \Phi_{(\centerdot, K+1)}\rangle$ and rearranging we obtain $\langle Y, \Phi_{(\centerdot, K+1)}\rangle=\sum_{n=1}^N X_n \langle \Phi_{(\centerdot, n)}, \Phi_{(\centerdot, K+1)}\rangle$. We can then write, 
\begin{dmath*}
\E[\langle Y,\Phi_{(\centerdot, K+1)}\rangle^2] =\E\left[\left(\sum_{n=1}^N X_n\langle \Phi_{(\centerdot, n)},\Phi_{(\centerdot, K+1)}\rangle\right)^2\right]\\
=\E\left[\sum_{n=1}^N (X_n)^2\langle \Phi_{(\centerdot, n)},\Phi_{(\centerdot, K+1)}\rangle^2\right]+\E\left[\sum_{n=1}^N\sum_{\substack{l=1 \\ l\neq n}}^N X_n X_l \langle \Phi_{(\centerdot, l)},\Phi_{(\centerdot, K+1)}\rangle \langle \Phi_{(\centerdot, n)}, \Phi_{(\centerdot, K+1)}\rangle\right]\\
=\sum_{n=1}^N \E[(X_n)^2]\E[\langle \Phi_{(\centerdot, n)},\Phi_{(\centerdot, K+1)}\rangle^2]+\sum_{n=1}^N\sum_{\substack{l=1 \\ l\neq n}}^N \E[X_n]\E[X_l]\E[\langle \Phi_{(\centerdot, l)},\Phi_{(\centerdot, K+1)}\rangle \langle \Phi_{(\centerdot, n)}, \Phi_{(\centerdot, K+1)}\rangle].
\end{dmath*}
The last step follows from the independence of $\Phi$ and $X$ and the independence of the $X_n$'s from each other. We claim that the cross-terms above sum to zero. To see this, we set $\Phi_{(\centerdot, l)}=a$, $\Phi_{(\centerdot, K+1)}=b$ and $\Phi_{(\centerdot, n)}=c$, where the entries of the vectors $a, b, c$ are all i.i.d. $\mathcal{N}(0,1)$. We note that if $l, K+1,$ and $n$ are mutually distinct, then $a, b, c$ are mutually independent. In this case we have, 
\begin{eqnarray*}
\E[\langle a, b\rangle \langle c, b \rangle] &=& \E[a^{\tpos} b c^{\tpos} b] \\
&=& \E[a^{\tpos}] \E[b c^{\tpos} b]\\
&=& 0.
\end{eqnarray*}

Since the cross-terms assume $l\neq n$, we consider the cases when either $n=K+1$ or $l=K+1$. In the case where $n=K+1$ we have, 
\begin{eqnarray*}
\E[\langle a, b\rangle \langle b, b \rangle] &=& \E[a^{\tpos} b b^{\tpos} b] \\
&=& \E[a^{\tpos}] \E[b b^{\tpos} b]\\
&=& 0.
\end{eqnarray*}

Similarly, in the case where $l=K+1$ we have, 
\begin{eqnarray*}
\E[\langle b, b\rangle \langle c, b \rangle] &=& \E[b^{\tpos} b c^{\tpos} b]\\
&=&\E[c^{\tpos} b b^{\tpos} b] \\
&=& \E[c^{\tpos}] \E[b b^{\tpos} b]\\
&=& 0.
\end{eqnarray*}

Thus, all cross-terms vanish so returning to our original goal we may claim,
\begin{dgroup*}
\begin{dmath*}
\E[\langle Y,\Phi_{(\centerdot,K+1)}\rangle^2]=\sum_{n=1}^N \E[(X_n)^2]\E[\langle \Phi_{(\centerdot, n)},\Phi_{(\centerdot, K+1)}\rangle^2]\\
= \sum_{n=1}^K\E[(X_n)^2]\E[\langle \Phi_{(\centerdot, n)},\Phi_{(\centerdot, K+1)}\rangle^2] + \E[(X_{K+1})^2]\E[\|\Phi_{(\centerdot, K+1)}\|^4] +\sum_{n=K+2}^N\E[(X_n)^2]\E[\langle \Phi_{(\centerdot, n)},\Phi_{(\centerdot, K+1)}\rangle^2].
\end{dmath*} 
\intertext{Examining each expected value individually, we recall that for $n \in \{1, \ldots, K\}=\anomalyset$ the $X_n$ were distributed with $\calD_2$ and thus $\E[(X_n)^2]= \E[X_n]^2+Var(X_n)= \mu_2^2+\sigma_2^2$. Recalling that the rest of the $X_n$ are distributed with $\calD_1$ which has $\mu=0$, we have that $\E[(X_n)^2]=\sigma_1^2$ in the subsequent cases. In~\cite{BWDSB05} they establish that $\E[\|\Phi_{(\centerdot, K+1)}\|^4]=M(M+2)$ and $\E[\langle \Phi_{(\centerdot, n)},\Phi_{(\centerdot, K+1)}\rangle^2]=M$, and we may use these results without further argument because we make the same assumptions about $\Phi$. Finally, substituting the expected values we have just calculated, we have that as $T$ grows large, the statistic $\xi_{n}$ when $n\notin \anomalyset$ converges to}
\begin{dmath*}
\E[\langle Y,\Phi_{(\centerdot, K+1)}\rangle^2] =K(\mu_2^2+\sigma_2^2)M+\sigma_1^2M(M+2)+(N-K-1)\sigma_1^2M\\
\end{dmath*}
\begin{dmath}
=M[K(\mu_2^2+\sigma_2^2)+(M+1+N-K)\sigma_1^2].
\label{eqn:osga1}
\end{dmath}
\end{dgroup*}

\runinhead{Anomalous Case:}
With $n=1$, we proceed as in the previous case,  
\begin{dgroup*}
\begin{dmath*}
\E[\langle Y,\Phi_{(\centerdot, 1)}\rangle^2]=\E\left[\left(\sum_{n=1}^N X_n\langle\Phi_{(\centerdot, n)},\Phi_{(\centerdot, 1)}\rangle\right)^2\right]\\
=\sum_{n=1}^N \E[(X_n)^2]\E[\langle \Phi_{(\centerdot, n)},\Phi_{(\centerdot, 1)}\rangle^2] ,\\
= \E[(X_1)^2]\E[\|\Phi_{(\centerdot, 1)}\|^4]+\sum_{n=2}^K\E[(X_n)^2]\E[\langle\Phi_{(\centerdot, n)},\Phi_{(\centerdot, 1)}\rangle^2]+\sum_{n=K+1}^N\E[(X_n)^2]\E[\langle\Phi_{(\centerdot, n)},\Phi_{(\centerdot, 1)}\rangle^2]\\
= (\mu_2^2+\sigma_2^2)M(M+2)+(K-2)(\mu_2^2+\sigma_2^2)M+(N-K)\sigma_1^2M\\
\end{dmath*}
\begin{dmath}
= M[(M+1+K)(\mu_2^2+\sigma_2^2)+(N-K)\sigma_1^2].
\label{eqn:osga2}
\end{dmath}
\end{dgroup*}

Combining the results of (\ref{eqn:osga1}) and (\ref{eqn:osga2}), we have
\[\lim_{T\rightarrow\infty}\xi_n=
\begin{cases}
M[(M+1+K)(\mu_2^2+\sigma_2^2)+(N-K)\sigma_1^2]& \quad n\in\anomalyset\\
M[K(\mu_2^2+\sigma_2^2)+(M+1+N-K)\sigma_1^2]&\quad n\notin\anomalyset.
\end{cases}
\]

The difference in the two expectations is thus,
\[
M(M+1)(\mu_2^2+\sigma_2^2-\sigma_1^2).
\]
For any $M\geq 1$ and $\mu_2^2+\sigma_2^2>\sigma_1^2$, the expected value of $\xi_n$ in the ``anomalous case'' is strictly larger than the expected value of $\xi_n$ in the ``prevalent case''. Therefore, as $T$ increases, OSGA can distinguish between the two expected values of $\xi_n$ with overwhelming probability. 
\qed
\end{proof}

The next theorem shows that asymptotically, Algorithm~\ref{alg:tecc} is able to detect anomalous variables with very few measurements at each time-step, for JSM-3R signals. Recall that JSM-3R signals have an unknown common component shared by signals at all time-steps, while each signal has a different innovation component that follows the JSM-2R model. The following theorem and proof assume that Algorithm~\ref{alg:osga} is implemented for step 6 of Algorithm~\ref{alg:tecc}. Once results like Theorem~\ref{thm:osga} exist for Algorithms \ref{alg:mmvsomp} and \ref{alg:mmvlasso}, then any JSM-2R algorithm could be used in step 6, and Theorem~\ref{thm:mmvtecc} would still hold. 

\begin{theorem}[Adapted from~\cite{BWDSB05} Theorem 10]
\label{thm:mmvtecc}
Let the $M\times N$ sensing matrix $\phi_t$ at each time-step $t$ contain entries that are i.i.d. $\sim\calN(0,1)$. For random variables $X_n$ that are distributed with $\calD_1=\mathcal{N}(\mu_1,\sigma_1^2)$ if $n\notin\anomalyset$ and $\calD_2=\mathcal{N}(\mu_2,\sigma_2^2)$ if $n\in\anomalyset$, if $\sigma_2^2>\sigma_1^2$ and with $M\geq 1$, TECC algorithm (with OSGA) recovers $\anomalyset$ with probability approaching one as $T\rightarrow \infty$.
\end{theorem}

We first note that the signals in Theorem {\ref{thm:mmvtecc}} correspond to the JSM-3R signals: for $n\notin\anomalyset$, the signal entries $x_{(n,t)}$ can be written as $x_{(n,t)}=\mu_1+x_{n,t}^I$ where $x_{n,t}^I$ are i.i.d. $\sim\calN(0,\sigma_1^2)$. With zero-mean and a potentially small variance, the amplitude of $x_{n,t}^I$, $n\notin\anomalyset$ is expected to be small. For $n\in\anomalyset$, the signal entries $x_{(n,t)}$ can be written as $x_{(n,t)}=\mu_2+x_{n,t}^I$ where $x_{n,t}^I$ are i.i.d. $\sim\calN(0,\sigma_2^2)$. With a larger variance $\sigma_2^2$, the amplitude of $x_{n,t}^I, n\in\anomalyset$ is expected to be much larger.

\begin{proof}
\smartqed
By the common component estimation from Algorithm~\ref{alg:tecc}, we have:
\begin{align*}
\widehat{x^C}&=\frac{1}{TM}\phi^{\tpos}y\\
&= \frac{1}{M}\frac{1}{T}\sum_{t=1}^T\phi_t^{\tpos}y_t\\
&= \frac{1}{M}\left(\frac{1}{T}\sum_{t=1}^T\phi_t^{\tpos} \phi_t \xdott\right).
\end{align*} 
Note that this is $1/M$ times the sample mean of the random variable $\Phi^{\tpos}\Phi X$. Letting $I_N$ denote the $N\times N$ identity matrix, we note that since $\Phi$ has independent $\calN(0,1)$ entries then $\E[\Phi^{\tpos}\Phi]=M I_N$. Since $\Phi$ is fully independent of $X$,
\[
\frac{1}{M}\E[\Phi^{\tpos}\Phi X]=\frac{1}{M}\E[\Phi^{\tpos}\Phi]\E[X] = I_N \E[X]=\E[X].
\]
Invoking the Law of Large Numbers, we have
\[
\lim_{T\rightarrow\infty}\widehat{x^C}=\E[X].
\]

Let $\widehat{X}=X-\widehat{x^C}$, then as $T\rightarrow\infty$, $\widehat{X}_n$ is distributed as $\mathcal{N}(0,\sigma_1^2)$ if $n\notin\anomalyset$ and $\mathcal{N}(0,\sigma_2^2)$ if $n\in\anomalyset$. Since $\widehat{Y}=Y-\Phi\widehat{x^C}=\Phi(X-\widehat{x^C})=\Phi\widehat{X}$, it follows from Theorem \ref{thm:osga} that with $M\geq 1$ and $\sigma_2^2>\sigma_1^2$, the TECC with OSGA algorithm recovers $\anomalyset$ with probability approaching one as $T\rightarrow\infty$.
\qed
\end{proof}

\section{Experiments}
\label{sec:experiments}
In this section, we evaluate numerically the performance of Algorithms \ref{alg:osga}, \ref{alg:mmvsomp}, \ref{alg:mmvlasso}, \ref{alg:tecc} and \ref{alg:acie} for anomaly detection. More specifically, we examine the success rate of determining the anomalous index set $\anomalyset$ from the signal matrix $x\in\mathbb{R}^{N\times T}$, whose columns are signals obtained at each time-step and share the same anomalous indices. The performance is assessed under various settings, by varying the number of anomalies, the number of columns in $x$ (i.e. the time-steps) and the number of mixed measurement $M$ at each time-step. Our focus is on the trade-off between the number of measurements $M$ and the number of time-steps $T$ required to identify $\anomalyset$ for varying numbers of anomalies.

In all experiments, the measurement matrices $\phi_t\in \R^{M\times N}$ comprise independent, $\calN(0,1)$ entries and the measurement vectors $y_t\in \R^M$ are calculated by $y_t=\phi_t \xdott$ for $t=1,\ldots,T$. To obtain an estimate of an algorithm's recovery success rate with high confidence, instead of using a fixed number of random trials across the different parameter combinations, we adaptively determine the necessary number of trials with a Jeffreys interval, a Bayesian two-tailed binomial proportion confidence interval. When the $95\%$ confidence interval around the true success rate shrinks to a width smaller than 0.1, we report the current proportion of successes as the recovery accuracy for the algorithm. The signals (i.e. $\xdott$) are generated under two models corresponding to the JSM-2R and JSM-3R signal definitions introduced in Section \ref{sec:method}. Algorithms \ref{alg:osga}, \ref{alg:mmvsomp} and \ref{alg:mmvlasso} are applied to the JSM-2R signals while Algorithms \ref{alg:tecc} and \ref{alg:acie} are applied to the JSM-3R signals.

The experiments are summarized in Table \ref{tab:exp}. The JSM-2R experiments assume a mean zero for the prevalent distribution and a much larger mean for the anomalous distribution while letting the variance be small. As shown in the previous section, signals generated from these distributions satisfy the definitions of JSM-2R. For JSM-3R experiments, we explore two settings: First, the prevalent and anomalous distributions are assumed to have different means; second, the two distributions have the same mean. Recall from the previous section, we show that the means of the distributions are the common components for the JSM-3 signals generated from these distributions. Note that the algorithms for JSM-3R signals have no knowledge of the mean of the prevalent or anomalous distributions.

\begin{table}[h!]
\centering
\caption{}
\label{tab:exp}
\begin{tabular}{|c|c|c|c|}
\hline
\textbf{Signal model}       & $\calD_1$& $\calD_2$& \textbf{Algorithms}  \\ \hline
JSM-2R & $\calN(0,1)$    & $\calN(7,1)$  & \begin{tabular}{@{}c@{}}OSGA, MMV-SOMP, \\ MMV-LASSO  \end{tabular}
  \\ \hline
JSM-3R & $\calN(7,1)$   & $\calN(0,10)$ & TECC, ACIE\\ \hline
JSM-3R & $\calN(7,1)$   & $\calN(7,10)$ & TECC, ACIE\\ \hline
\end{tabular}
\end{table}

We chose the distributions in Table~\ref{tab:exp} for our numerical simulations to remain consistent with~\cite{CXL16}. We observe, in the JSM-2R experiments, that the distributions $\calN(0,1)$ and $\calN(7,1)$ have their means separated by three standard deviations each, with one additional standard deviation in between for good measure. This ensures that the distributions are statistically distinct from each other. We have not explored how the detection accuracy is affected as we vary the proportion of overlap in the two distributions.

\subsection{JSM-2R}
\label{subsec:jsm2r}
\begin{figure}[h!]
\includegraphics[width=\textwidth]{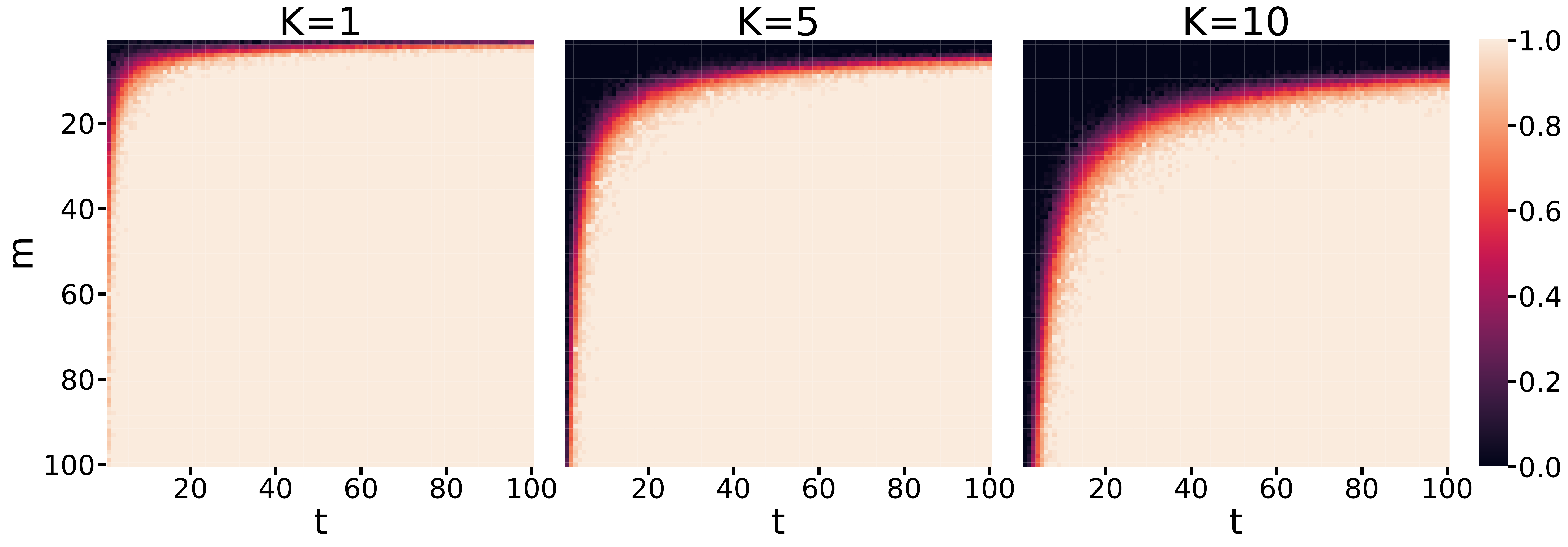}
\caption[OSGA Recovery Phase Transition]{The recovery phase transition for the OSGA algorithm with $K=1,$  5, and 10 anomalous random variables.}
\label{fig:mmvosgatriple}
\end{figure}

\begin{figure}[h!]
\includegraphics[width=\textwidth]{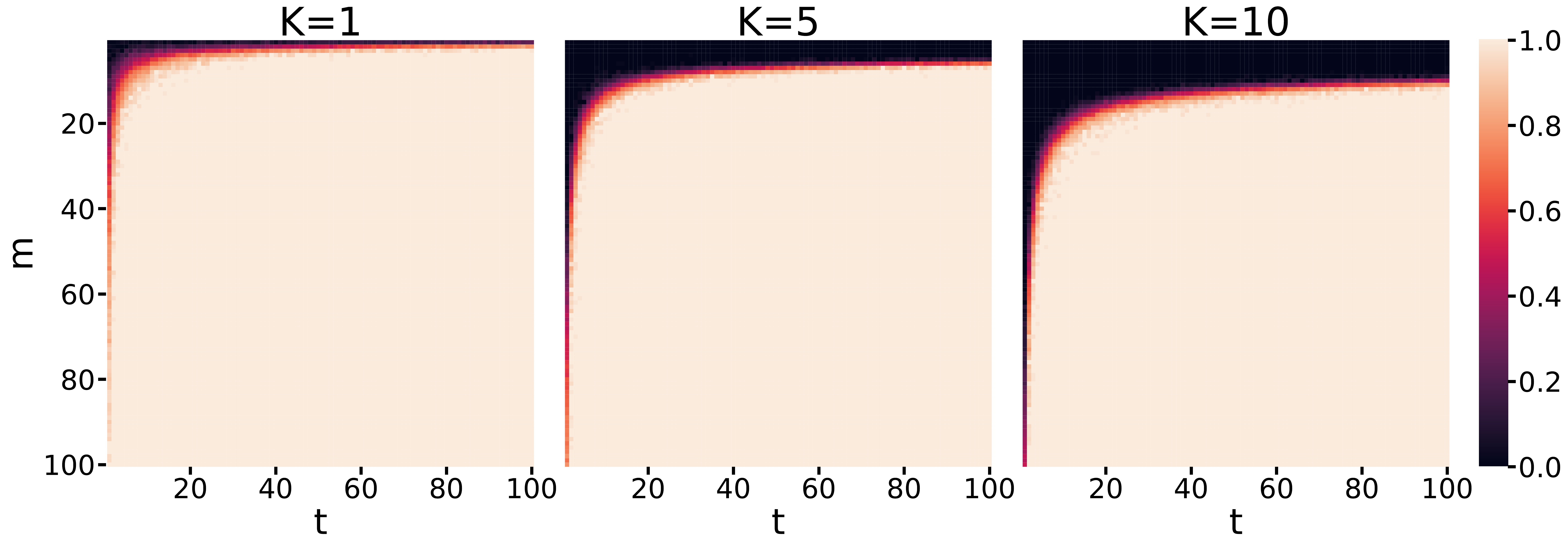}
\caption[MMV-SOMP Recovery Phase Transition]{The recovery phase transition for the MMV-SOMP algorithm with $K=1,$ 5, and 10 anomalous random variables.}
\label{fig:mmvsomptriple}
\end{figure}

\begin{figure}[h!]
\includegraphics[width=\textwidth]{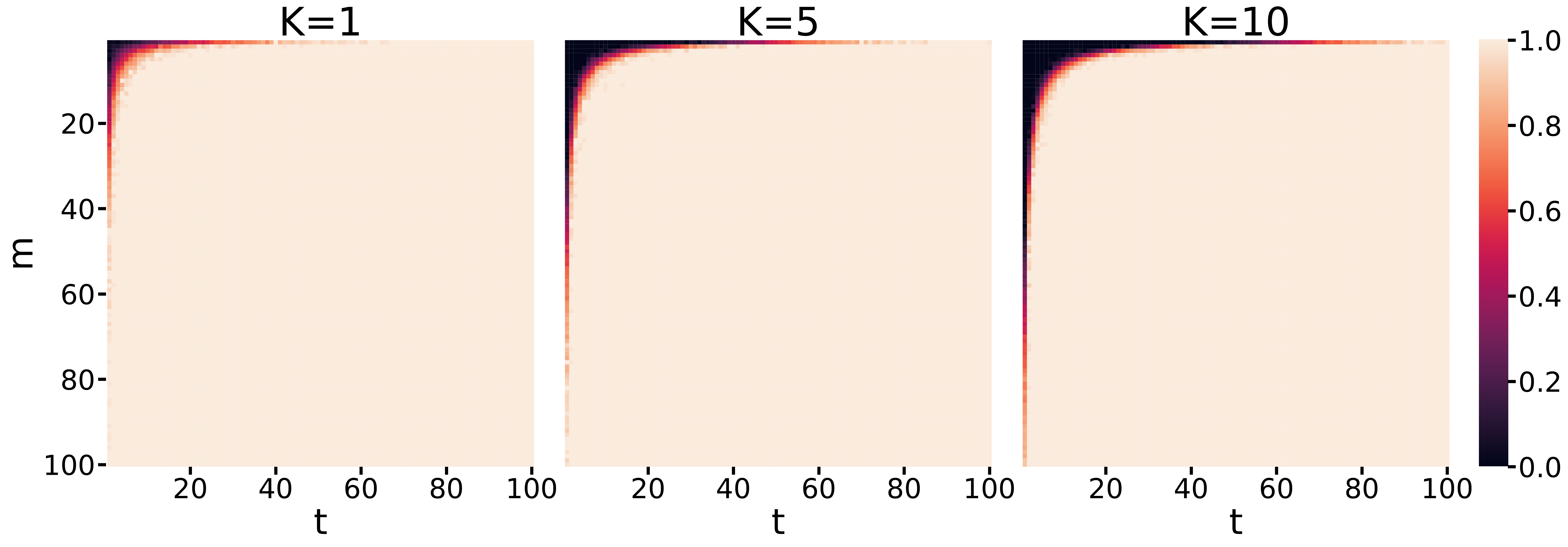}
\caption[MMV-LASSO Recovery Phase Transition]{The recovery phase transition for the MMV-LASSO algorithm with $K=1,$ 5, and 10 anomalous random variables.}
\label{fig:mmvlassotriple}
\end{figure}

We now present the results of recovering the anomalous index set for the JSM-2R signals. The signal length is fixed at $N=100$ and results of $K=1, 5,$ and $10$ anomalies are presented. For each $K$ value, $K$ random variables follow the distribution $\calN(7,1)$ and the other $N-K$ random variables follow another distribution $\calN(0,1)$. The goal is to recover the index set $\anomalyset$ of these $K$ random variables. Figure~\ref{fig:mmvosgatriple} shows the success rate of identifying $\anomalyset$ for the three $K$ values using the OSGA Algorithm. Each dot in the figure denotes the success rate for a specific $M$ (number of measurements per time-step) and a specific $T$ (number of time-steps) estimated from a number of trials, and the value is indicated by the color (see the colorbar). Both $M$ and $T$ take values from $1$ to $100$. Figure~\ref{fig:mmvsomptriple} and~\ref{fig:mmvlassotriple} plot the success rate for MMV-SOMP and MMV-LASSO algorithms respectively.

For all three algorithms, the success rate of anomaly identification increases as the number of measurements $M$ increases and/or as the number of time-steps $T$ increases. A 100\% success of identification is obtained with a sufficiently large number of measurements and time-steps. There are some important differences in performance among the three algorithms.

Firstly, for the OSGA and the MMV-SOMP algorithms, with a sufficiently large number of time-steps, the minimum number of measurements at each time-step required for anomaly detection increases with the number of anomalies present. The MMV-LASSO performance seems less affected by varying the number of anomalies than the performance of the other two algorithms. Secondly, comparing Figure~\ref{fig:mmvosgatriple} and~\ref{fig:mmvsomptriple} reveals that MMV-SOMP requires fewer time-steps than the OSGA algorithm to reach 100\% success for a given number of measurements. Thirdly, the MMV-LASSO algorithm requires significantly fewer measurements and time-steps for 100\% success compared with OSGA and MMV-SOMP. Finally, there is asymmetry between the effect of increasing the number of measurements versus that of increasing the number of time-steps on the performance of OSGA and MMV-SOMP. For these two algorithms, increasing the number of measurements is more effective than increasing the number of time-steps for improving the performance. No obvious asymmetry of recovery performance is found for the MMV-LASSO algorithm. The near symmetry phenomenon of the MMV-LASSO is expected since doubling either $M$ or $T$ doubles the number of rows in the matrix $\phi$ in Algorithm \ref{alg:mmvlasso}, providing similar amounts of information for the algorithm. 

For comparison with a benchmark, we note that in~\cite{CXL16}, the authors propose LASSO as an efficient algorithm to detect anomalies. The performance of their proposed method is shown as the first row, $M=1$, in the phase diagrams of Figure~\ref{fig:mmvlassotriple}. Here, we expand the application of LASSO by allowing for a trade-off between the number of measurements per time-step, $M$, and the number of time-steps, $T$, for which measurements are taken. Applications with an ability to store multiple measurements at each time-step, while seeking to minimize the time needed to accumulate data, might prefer to use the MMV-LASSO of Algorithm~\ref{alg:mmvlasso} to detect anomalies.

\begin{figure}[h!]
\centering
\includegraphics[width=5.8cm]{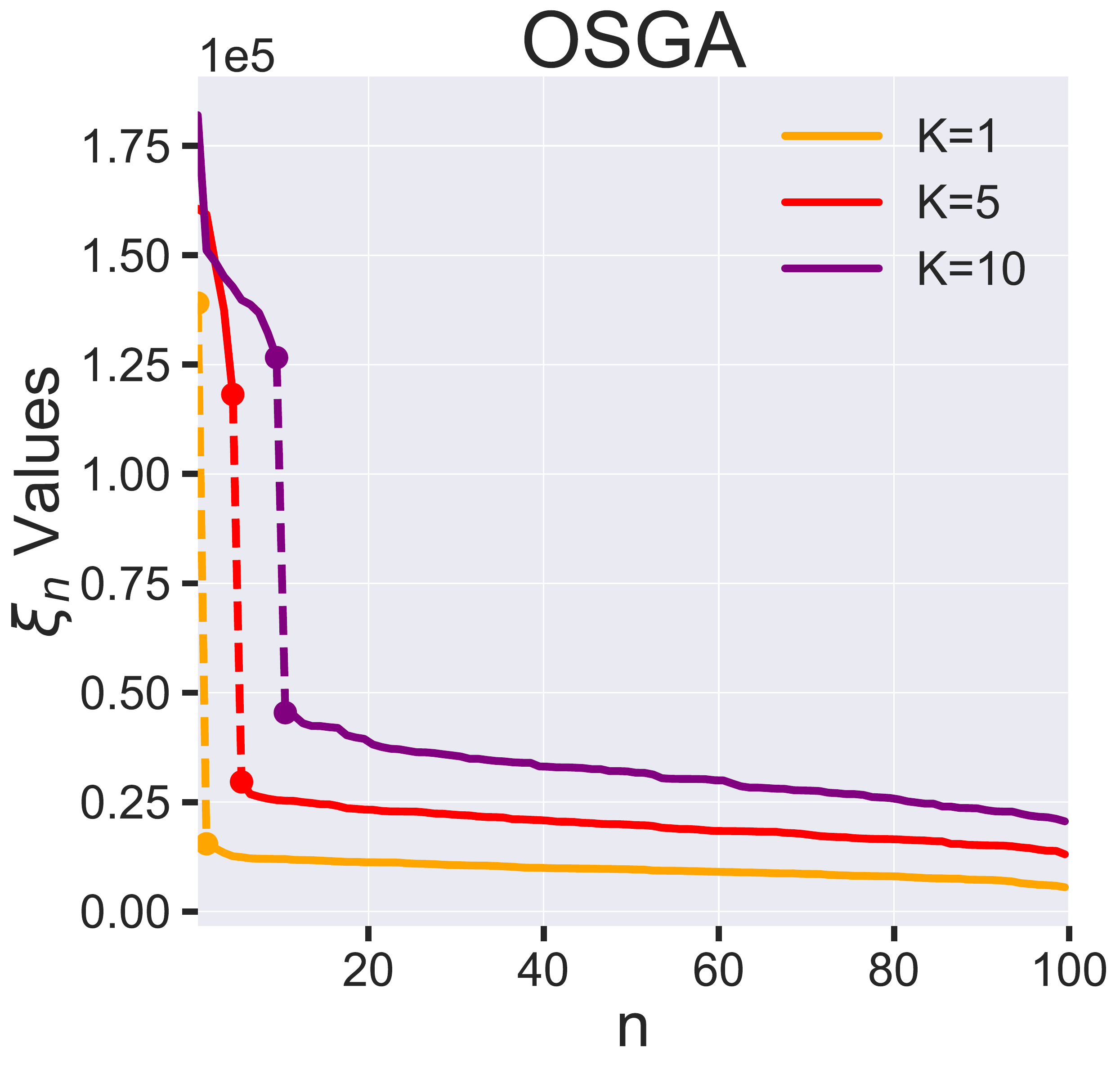}\\
\includegraphics[width=5.8cm]{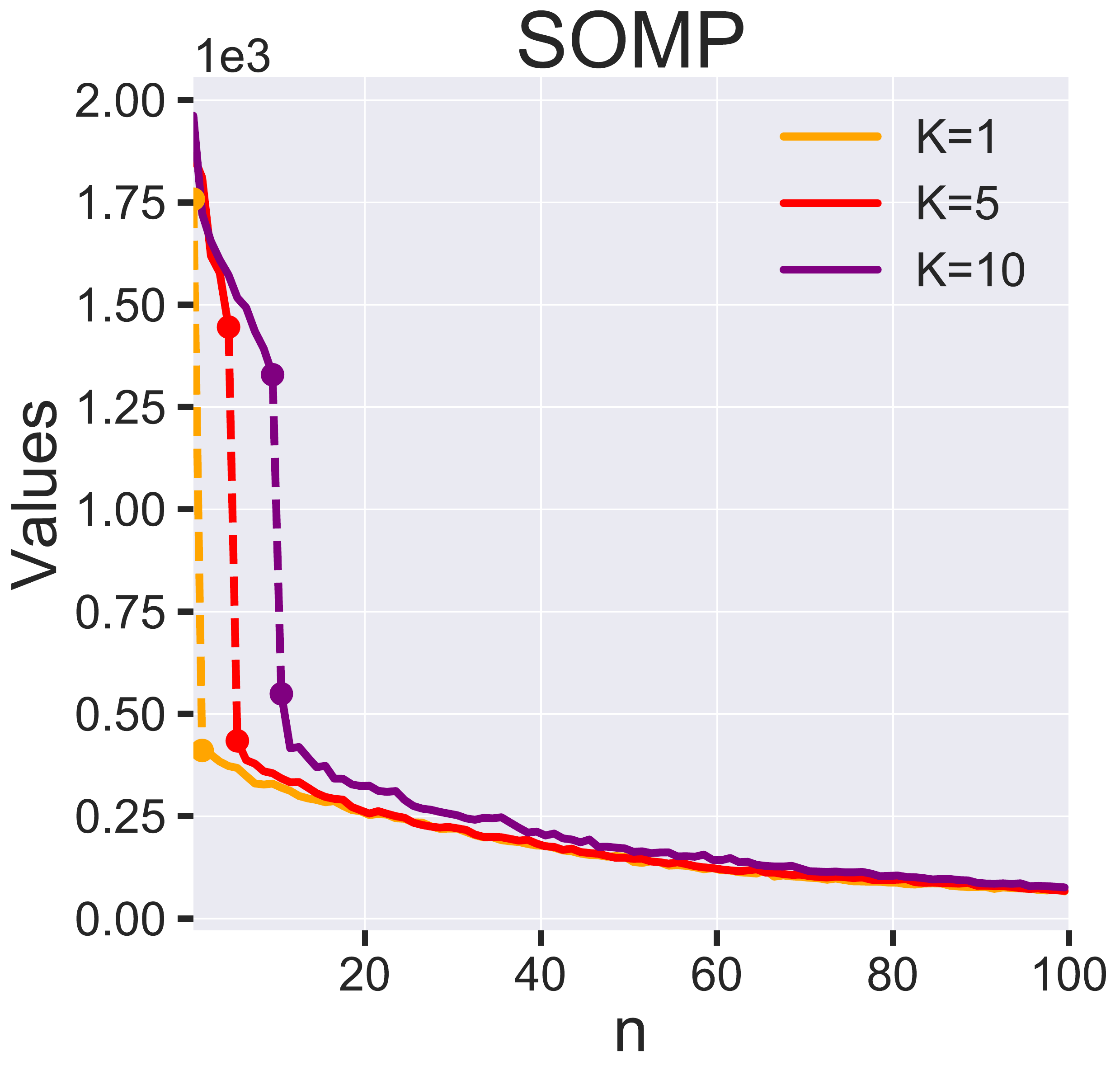}\\
\includegraphics[width=5.8cm]{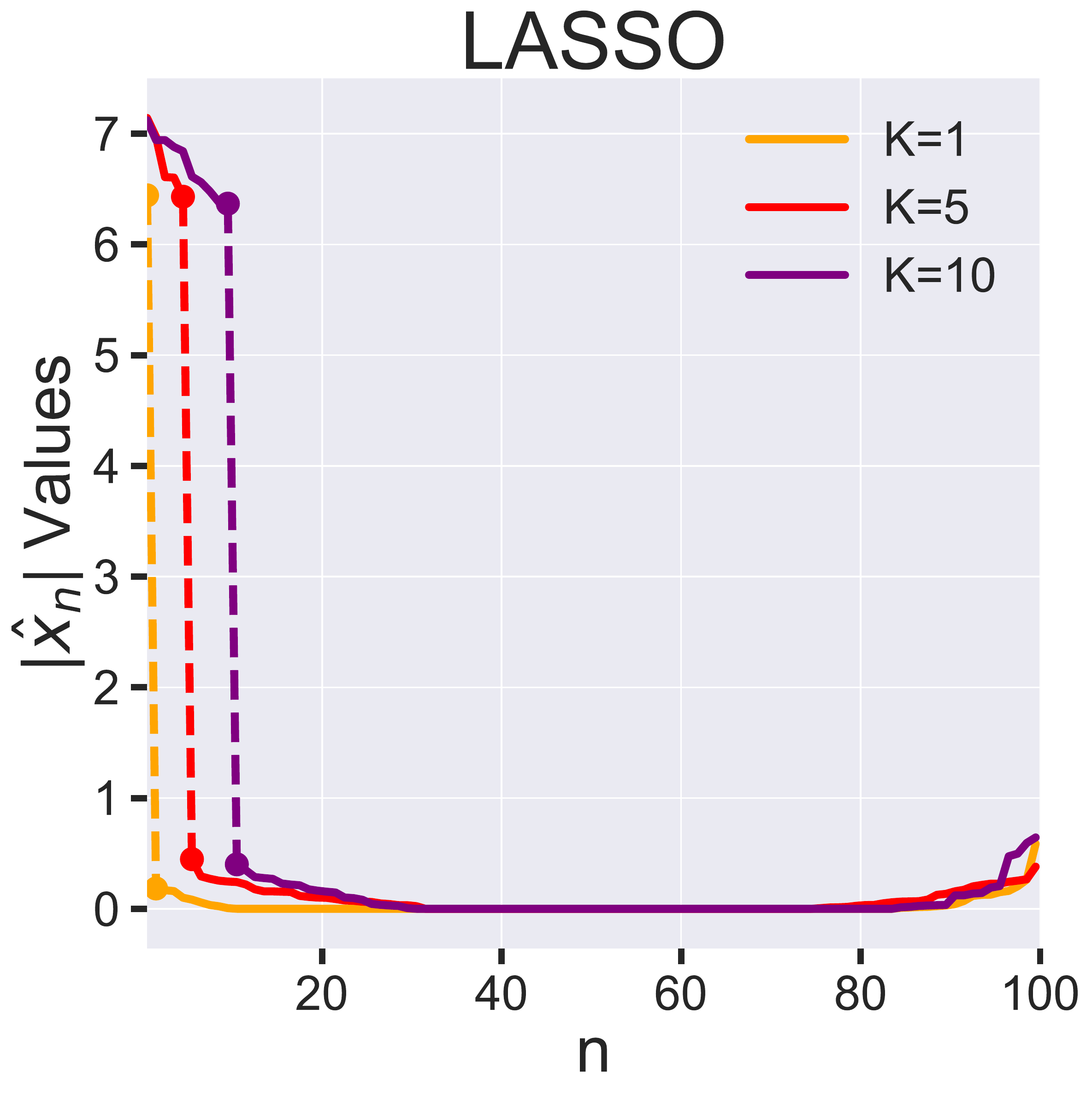}
\caption[Inferred $K$: OSGA]{Plots of the values from which indices are selected for $\widehat{\anomalyset}$ in the JSM-2R algorithms. The dotted line denotes the drop between the top $K$ values and the remaining $N-K$ values.}
\label{fig:noktriple}
\end{figure}

In these experiments, we have assumed that we know the number of anomalies, $K$. To explore the possibility of estimating the number of anomalies as we detect them, we consider the following experiments.
\begin{enumerate}
\item For OSGA, we calculate the test statistics, $\xi_n$, in Algorithm~\ref{alg:osga} for all $n=1,2,\cdots,N$ and sort them in descending order; then determine whether the amplitudes of the $\xi_n$ can be used to estimate $K$.
\item Similarly, for MMV-SOMP, we use the amplitude of $\sum_{t=1}^T\frac{|<r_t^{k-1},\phi_{t(\cdot,n)}>|}{\|\phi_{t(\cdot,n)}\|_2}$ in Algorithm~\ref{alg:mmvsomp} to determine the number of anomalies.
\item Lastly, for MMV-LASSO, we calculate the reconstructed signal $|\hat{x}|$ in Algorithm~\ref{alg:mmvlasso} and sort the entries in descending order, and determine $K$ based on the amplitudes.
\end{enumerate}
In each case, we fix $M=50$ and $T=50$ to ensure that recovery is possible if $K$ is known (we can see this from the results in Figures~{\ref{fig:mmvosgatriple}}, {\ref{fig:mmvsomptriple}},{\ref{fig:mmvlassotriple}}). The results shown in Figure~{\ref{fig:noktriple}} demonstrate the potential of these methods to estimate $K$. Theoretical justification of these methods is left as future work.

\subsection{JSM-3R}
\begin{figure}[h!]
\includegraphics[width=\textwidth]{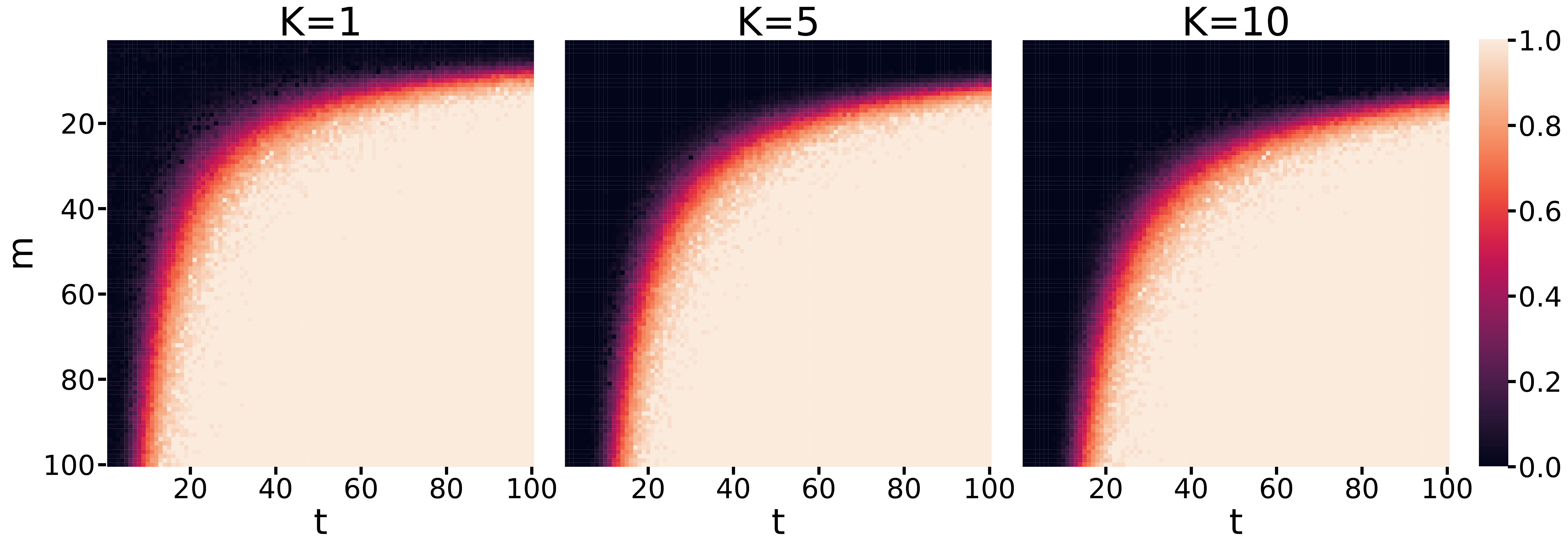}
\caption[TECC Recovery Phase Transition]{The recovery phase transition for the TECC algorithm with $K=1,$  5, and 10 anomalous random variables. Here the prevalent distribution is $\calN(7,1)$ and the anomalous distribution is $\calN(0, 10)$.}
\label{fig:tecctriple}
\end{figure}

\begin{figure}[h!]
\includegraphics[width=\textwidth]{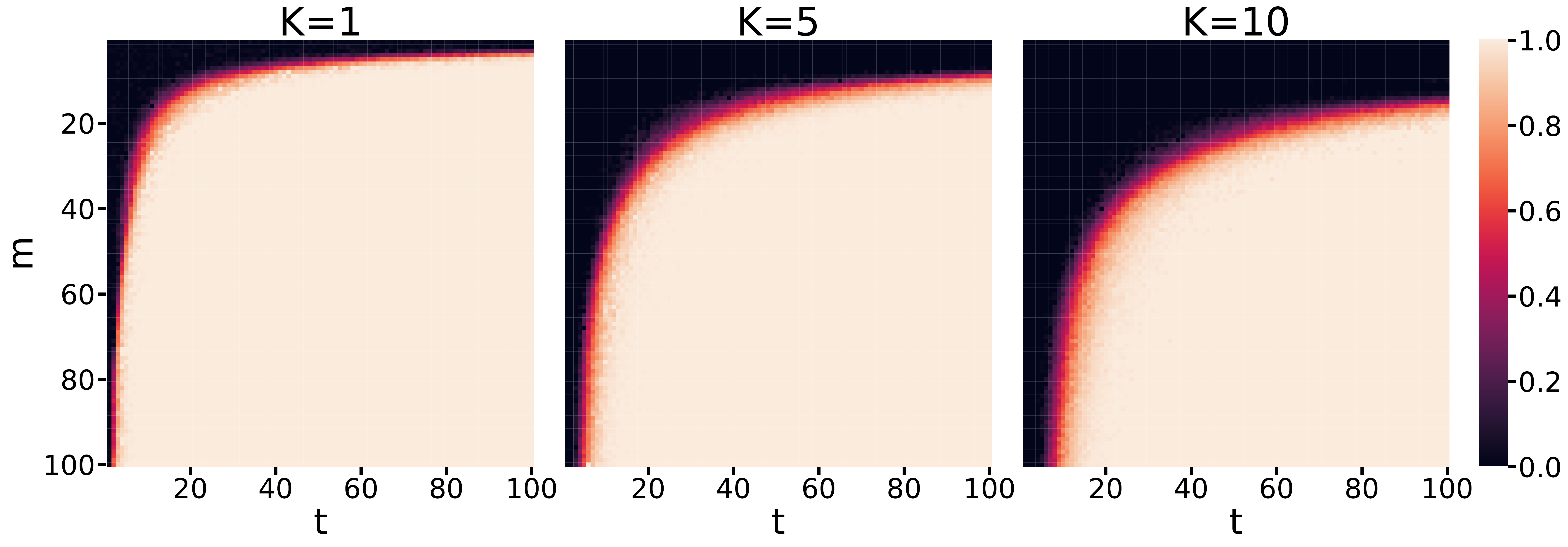}
\caption[ACIE Recovery Phase Transition]{The recovery phase transition for the ACIE algorithm with $K=1$, 5, and 10 anomalous random variables. Here the prevalent distribution is $\calN(7,1)$ and the anomalous distribution is $\calN(0, 10)$.}
\label{fig:acietriple}
\end{figure}

We next present the results of recovering the anomalous index set for the JSM-3R signals. Similar to JSM-2R signals, the length of the signal is set to $N=100$ and the number of anomalies takes values of $K=1,5,$ and $10$. Unlike the JSM-2R signals, the $N-K$ random variables now follow the distribution $\calN(7,1)$ while the $K$ anomalous random variables follow the distribution $\calN(0,10)$ or $\calN(7,10)$. In order for a fair comparison between the algorithms, we implement the OSGA algorithm for both step $6$ of the TECC algorithm and step $10$ of the ACIE algorithm. The iteration $L$ in the ACIE algorithm is set to $L=5$.  The performance of the TECC and ACIE algorithms for varying numbers of measurements $M$ and time-steps $T$ when the anomalous distribution follows $\calN(0,10)$ is presented in Figures~\ref{fig:tecctriple} and ~\ref{fig:acietriple}, where both $M$ and $T$ range from $1$ to $100$. The performance for the setting where the anomalous variables are distributed as $\calN(7,10)$ is similar to Figures~\ref{fig:tecctriple} and~\ref{fig:acietriple} and is thus omitted.

With a sufficiently large number of measurements and time-steps, both algorithms are able to achieve 100\% success in recovery of the anomalous index set. For a fixed number of time-steps, the minimum number of measurements required for identification increases as the number of anomalies increases for both algorithms. There is improvement in performance of the ACIE algorithm over the TECC algorithm. The ACIE algorithm requires fewer time-steps to reach 100\% recovery success, for a given number of measurements; similarly, it requires fewer measurements for 100\% recovery success with a given number of time-steps. 

\begin{figure}[h!]
\includegraphics[width=\textwidth]{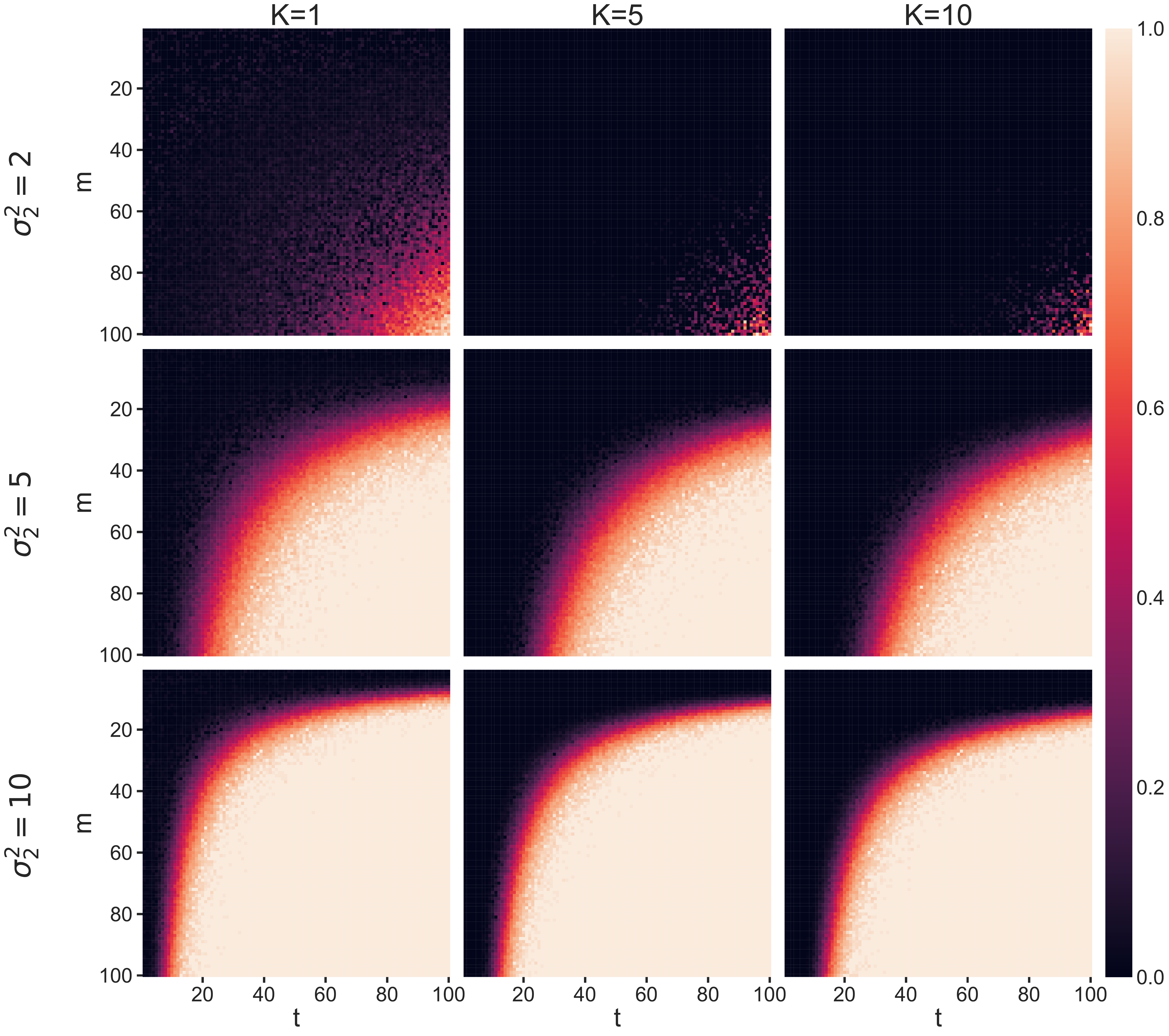}
\caption[Varying $\sigma_2^2/\sigma_1^2$]{The recovery phase transition for the TECC algorithm with $K=1$, 5 and 10 anomalous random variables. Here the prevalent distribution is $\calN(7,1)$ and the anomalous distribution is $\calN(0, \sigma_2^2)$, with $\sigma_2^2=2$, 5,  and 10 shown.}
\label{fig:teccsigma}
\end{figure}

\begin{figure}[h!]
\includegraphics[width=\textwidth]{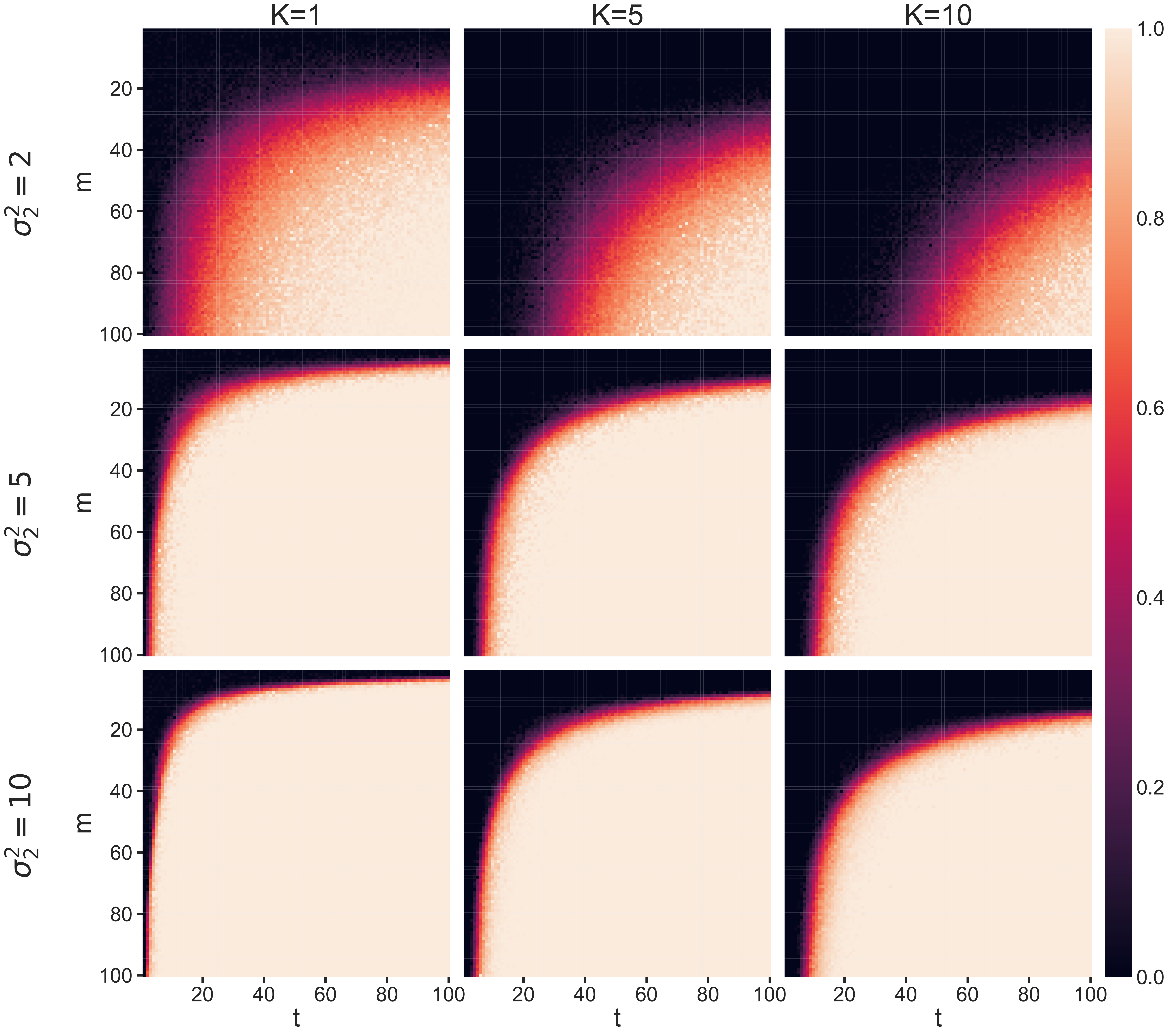}
\caption[Varying $\sigma_2^2/\sigma_1^2$]{The recovery phase transition for the ACIE algorithm with $K=1$, 5 and 10 anomalous random variables. Here the prevalent distribution is $\calN(7,1)$ and the anomalous distribution is $\calN(0, \sigma_2^2)$, with $\sigma_2^2=2$, 5,  and 10 shown.}
\label{fig:aciesigma}
\end{figure}

Thus far, we have assumed that the prevalent and anomalous distributions have very different variances, $\sigma_1^2=1$ and $\sigma_2^2=10$ in these experiments. To investigate the performance of these algorithms as the ratio of the variance changes, we experiment by setting $\sigma_2^2/\sigma_1^2=2, 5,$  and $10$, for $K=1, 5,$ and $10$. Figure~{\ref{fig:teccsigma}} shows the phase transition for the TECC algorithm as we vary the ratio of the variances, and Figure~{\ref{fig:aciesigma}} shows the phase transition for the ACIE algorithm as we vary the ratio of the variances. In both cases, the algorithms are behaving as we might expect. The smaller the ratio between the variances, the more measurements and time-steps it takes to detect the anomalies.

\section{Conclusion}
\label{sec:conclusion}
In this paper, we formally posed the problem of detecting anomalously distributed random variables as an MMV problem, by drawing an analogy between samples of the random variables and ensembles of signals. We further established two signal models characterizing possible correlation structures among signals that contain anomalous entries. Based on the new signal models, we showed through theoretical and numerical analysis that many of the MMV algorithms for sparse signal recovery can be adapted to the anomaly detection problem. For two of the algorithms, we provided theoretical guarantees of anomaly detection in the asymptotic case. Our experimental results on synthetic data show good performance for signals conforming to either model, when a sufficiently large number of time-steps is available. 

While these algorithms succeed in detecting anomalies, there is still room for optimizing performance. Currently these algorithms require storing the sensing matrices at each time-step in memory. In future work, we would like to explore optimal ways to design sensing matrices to reduce the memory burden. Having provided asymptotic anomaly detection guarantees for two algorithms,  we are further interested in providing such guarantees for all the algorithms presented. Additionally, we are interested in characterizing the performance bounds for each algorithm's finite sample case. Theorem~{\ref{thm:mmvtecc}} shows that only when the variances of the anomalous and prevalent distributions are distinct can the anomalies be detected by the algorithm. With additional information about the means of the distributions, perhaps the algorithms could be extended to identify the differences in means and detect anomalies even with identical variances. Finally, the theoretical results presented rely on Gaussian distributions. We are interested in expanding these algorithms to distributions which might not be distinguishable with the current approach. For distributions with heavy tails where the variance is no longer finite, a theorem assuming the law of large numbers might be incorrect, or the convergence to the expected value might be very slow. It would be interesting to investigate under what kinds of heavy-tailed distributions these algorithms start to fail.

\begin{acknowledgement}
The initial  research for this effort was conducted at the Research Collaboration Workshop for Women in Data Science and Mathematics, July 17-21 held at ICERM. Funding for the workshop was provided by ICERM, AWM and DIMACS (NSF grant CCF-1144502). SL was supported by NSF CAREER grant CCF$-1149225$. DN was partially supported by the Alfred P. Sloan Foundation, NSF CAREER $\#1348721$, and NSF BIGDATA $\#1740325$. JQ was supported by the faculty start-up fund of Montana State University. 
\end{acknowledgement}

%
\pagebreak
\section*{Appendix}
\label{sec:notation}
\addcontentsline{toc}{section}{Appendix}
Here we provide a summary of notation for reference. 
\begin{center}
\begin{tabular}{r|l}
$N$ & Number of random variables\\
$\indexset$ & Set of random variables indices, $\{n\in \mathbb{N}: 1\leq n \leq N\}$\\
$K$ & Number of anomalous random variables\\
$\anomalyset$ & Set of anomalous random variable indices, $\anomalyset\subset\indexset$, $|\anomalyset| = K$\\
$M$ & Number of measurements per time-step\\
$m$ & Measurement index, $1\leq m\leq M$\\
$T$ & Number of time-steps measured\\
$t$ & Time-step index, $1\leq t \leq T$\\
$\calD_1$ & Prevalent distribution \\
$\calD_2$ & Anomalous distribution \\
$X$ & Random vector comprising independent random variables $X_1, \ldots, X_N$\\
$x$ & $N\times T$-dimensional matrix of independent realizations of $X$ for all $T$ time-steps\\
$\Phi$ & $M\times N$-dimensional sensing matrix, i.i.d. $\sim\calN(0,1)$ entries \\
$\phi_t$ & $M\times N$-dimensional realization of $\Phi$ at time $t$ \\
$\phi$ & $(M\cdot T)\times N$-dimensional vertical concatenation of the $\phi_t$, $[\phi_1^{\tpos},\ldots, \phi_T^{\tpos} ]^{\tpos}$\\
$y_{t}$ &  $M$-dimensional result of measuring the signal, $\phi_t\cdot\xdott$, at time $t$\\
$y$ & $(M\cdot T)$-dimensional vertical concatenation of the $y_t$, $[y_1^{\tpos}, \ldots, y_T^{\tpos}]^{\tpos}$\\
$Y$ & $M$-dimensional random vector defined by $\Phi X$ \\ 
JSM & Joint Sparsity Model, introduced in~\cite{BWDSB05}\\
JSM-2 & Signals are nonzero only on a common set of indices\\
JSM-3 & Signals consist of common non-sparse component and a sparse innovation\\
JSM-2R & ``Random variable'' version of JSM-2\\
JSM-3R & ``Random variable'' version of JSM-3\\
OSGA & One-step greedy algorithm\\
MMV & Multiple Measurement Vector\\
MMV-LASSO & MMV Least Absolute Shrinkage and Selection Operator\\
MMV-SOMP & MMV Simultaneous Orthogonal Matching Pursuit\\
TECC & Transpose Estimation of Common Component\\
ACIE & Alternating Common and Innovation Estimation\\ 
\end{tabular}
\end{center}
We adopt the convention that random variables will be upper case and their realizations will be lower case. All matrix entries will have two, subscripted indices. The first index will indicate the row position, the second will indicate the column position. We will indicate row and column vectors by substituting $\centerdot$ for the respective index.

\pagebreak
\bibliographystyle{spbasic}
\bibliography{icerm_references}

\end{document}